\documentclass[lettersize,journal]{IEEEtran}
\usepackage{amsmath,amsfonts}
\usepackage[linesnumbered,ruled]{algorithm2e}
\usepackage{array}
\usepackage[caption=false,font=normalsize,labelfont=sf,textfont=sf]{subfig}
\usepackage{textcomp}
\usepackage{stfloats}
\usepackage{url}
\usepackage{verbatim}
\usepackage{graphicx}
\usepackage{cite}
\usepackage{amssymb}

\usepackage{epstopdf}
\newcommand{\PreserveBackslash}[1]{\let\temp=\\#1\let\\=\temp}
\newcolumntype{C}[1]{>{\PreserveBackslash\centering}p{#1}}
\newcolumntype{R}[1]{>{\PreserveBackslash\raggedleft}p{#1}}
\newcolumntype{L}[1]{>{\PreserveBackslash\raggedright}p{#1}}
\usepackage[usenames]{color}
\usepackage{colortbl,booktabs}
\usepackage{tabularx}
\usepackage{multicol}
\usepackage{booktabs}
\usepackage[Symbol]{upgreek}
\usepackage{bm}
\usepackage{balance}
\usepackage{mathrsfs}
\usepackage{threeparttable}
\usepackage{dcolumn}
\usepackage{multirow}
\usepackage{booktabs}
\usepackage{float}
\usepackage{soul}
\usepackage{xcolor}

\begin{document}
	\title{A Joint Graph-Cut Channel Estimation Method for Multi-user Holographic MIMO}
	
	\author{Jiaxin Zhang, Wenqian Shen,~\IEEEmembership{Member,~IEEE,} Kai Yang,~\IEEEmembership{Member,~IEEE,} Zhen Gao,~\IEEEmembership{Member,~IEEE,} and Jianping An,~\IEEEmembership{Senior Member,~IEEE}
		
	\thanks{
	Copyright (c) 2026 IEEE. Personal use of this material is permitted. However, permission to use this material for any other purposes must be obtained from the IEEE by sending a request to pubs-permissions@ieee.org. 
	
	This work was supported in part by National Natural Science Foundation of China under Grant No. 62571041, in part by the Natural Science Foundation of Beijing, China under Grant No. 4252010, in part by the National Key Research and Development Program of China under Grant No. 2024YFC3306801, and the National Key Laboratory of Science and Technology on Space Microwave under Grant HTKJ2024KL504006. (Corresponding author: Wenqian Shen.)
	
	J. Zhang, W. Shen, K. Yang, Z. Gao and J. An are with School of Information and Electronics, Beijing Institute of Technology, Beijing 100081, China (e-mail: bit\_ZhangJX@163.com; shenwq@bit.edu.cn; yangkai@bit.edu.cn; gaozhen16@bit.edu.cn; an@bit.edu.cn).}

	\vspace*{-5mm}	
	}

	\maketitle
	\begin{abstract}
		To address the challenges of high-dimensional channel estimation and underutilized spatial correlations among users in holographic MIMO (HMIMO) systems, this paper proposes a joint graph-cut algorithm for multi-user channel estimation in the wavenumber domain.
		The size of the conventional angular-domain channel matrix increases with the number of antennas in densely-spaced HMIMO. Therefore, user channels are projected into the wavenumber domain via a Fourier harmonic transform, revealing their inherent clustered sparsity and exploiting common scatterer clusters among users. Subsequently, a joint graph-cut channel estimation (JGC-CE) algorithm based on multi-user common supports is designed. In each iteration, the algorithm first partitions user clusters to extract shared supports. Then for each user, it performs users' individual graph update and channel estimation to reconstruct the channel matrix. Simulation results demonstrate that the proposed method outperforms independent estimation schemes for individual users in accuracy while reducing pilot length.
	\end{abstract}
	
	\IEEEpeerreviewmaketitle
	
	\section{Introduction}\label{S1}
	
	Holographic MIMO (HMIMO), with its near-continuous aperture and superior electromagnetic wave manipulation capabilities, has emerged as a pivotal enabling technology for 6G\cite{1,6}.
	Thanks to the development of hypersurface material technology, some subwavelength-scale antenna array HMIMO  architectures have emerged \cite{11,18}, offering an exceedingly high degree of freedom (DoF) in electromagnetic manipulation.
	HMIMO can enable beamforming with super-directivity \cite{16}. To fully achieve the beneficial beamforming capabilities in HMIMO system, it is imperative to develop an efficient and reliable channel state information (CSI) acquisition method. 
	
	Recent studies have explored HMIMO channel modeling and estimation for single-user scenarios. Prior work \cite{2} investigated least squares (LS) and minimum mean square error (MMSE) estimation under far-field assumptions, but these conventional methods failed to handle HMIMO's high-dimensional channel matrices.
	To address this, \cite{3} employed Fourier plane-wave series expansions to characterize HMIMO's electromagnetic propagation, introducing the concept of wavenumber domain, which serves as a natural tool for describing continuous and quasi-continuous aperture electromagnetic fields and is valid in both far-field and near-field. The dimension of the wavenumber-domain channel is much lower than that of the angular-domain. Following this insight, \cite{4} adapted wavenumber-domain orthogonal matching pursuit (WD-OMP) algorithm for channel estimation by leveraging the sparsity, while \cite{5,12} introduced a Markov random field (MRF) model to capture clustered sparsity, improving estimation via graph partitioning. 
	
	A critical limitation persists in existing HMIMO research: nearly all published works focus on single-user cases, leaving multi-user channel estimation unexplored. Using orthogonal pilot signals to extend the single-user method to multi-user leads to excessively high pilot overhead.
	Notably, studies \cite{7,8} in traditional multi-user MIMO systems reveal that shared local scattering clusters often lead to common sparsity structures across user channel matrices. This insight can be extended to HMIMO: when users coexist in the same scattering environment, their channels exhibit similar wavenumber-domain sparsity patterns. The inherent correlation provides a foundation for generalizing HMIMO estimation to multi-user scenarios with low overhead.
	However, the common supports exhibited by HMIMO in the wavenumber domain do not exactly match the conclusions of mMIMO in the angular domain. This is due to the fact that different user channels have different weights of angular energy spectrum function\cite{12} on the wavenumber domain, leading to a more ambiguous and unrecognizable boundary of the common support set. 
	
	This work bridges the knowledge gap by exploiting inter-user shared scatterers to construct a joint sparse channel model, thereby enhancing estimation accuracy while reducing pilot resources and computational complexity. 
	Firstly, by drawing on group sparsity theory from conventional MIMO systems, the multi-user channel estimation problem can be reformulated as a structured sparse recovery problem. Secondly, to solve this problem, we propose a novel joint graph-cut channel estimation (JGC-CE) algorithm that combines  common support set extraction and graph-cut based individual user channel estimation methods. Finally, comprehensive simulations demonstrate the effectiveness of the proposed algorithm, showing significant improvements in both estimation accuracy and computational efficiency compared to conventional approaches.

	\section{System Model}\label{S2}

	\begin{figure}[t]
		\center{\includegraphics[width=0.4\textwidth]{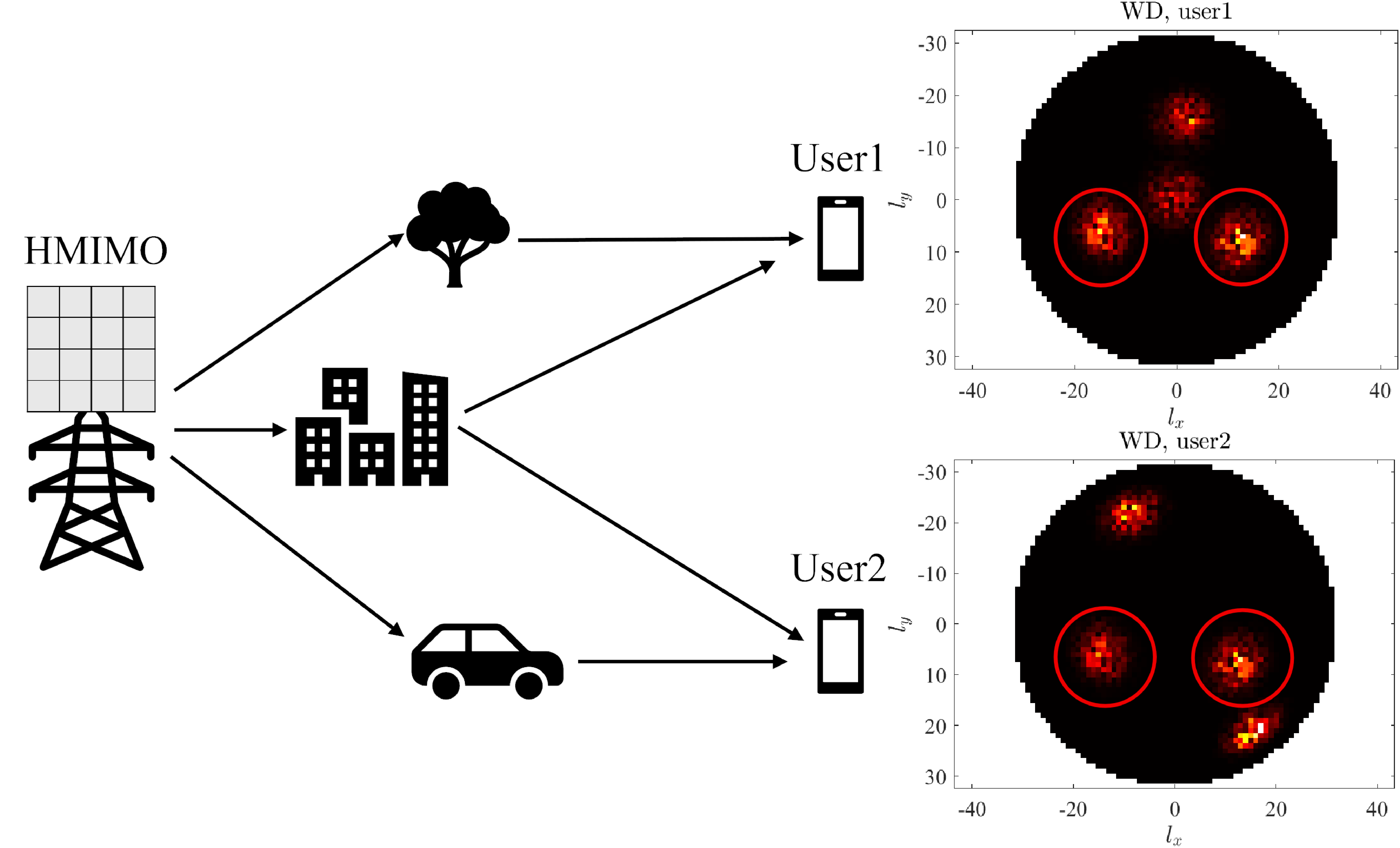}}
		\caption{Common support in multi-user HMIMO system.}
		\label{fig1}
	\end{figure}

	Consider a downlink HMIMO system operating at carrier frequency $f_c$, consisting of a base station (BS) and $K$ single-antenna users, as illustrated in Fig. \ref{fig1}. The BS transmitter is equipped with a uniform planar array (UPA) with element spacing $\delta$, where $\delta<\dfrac{\lambda}{2}$ (with $\lambda$ denoting the wavelength). The array size of the UPA is $L_{x} \times L_{y}$, and it comprises $N = N_{x} \times N_{y}$ antenna elements, where $L_{x} = (N_{x} - 1)\delta$, $L_{y} = (N_{y} - 1)\delta$. Let $\mathbf{x}_{j} \in \mathbb{C}^{N \times 1}$ denote the pilot signal transmitted by the BS in the	$j$-th time slot. Then, the corresponding received signal at the $i$-th user, denoted as ${y}_{ij}$, can be expressed as:
	\begin{align}\label{H_i}
		{y}_{ij} = \mathbf{x}^T_{j}\mathbf{h}_{i} + {n}_{ij},
	\end{align}
	where 
	$\mathbf{h}_{i} \in \mathbb{C}^{N \times 1}$ is the channel vector between the BS and the $i$-th user, and ${n}_{ij}$ denotes the additive complex Gaussian noise at the receiver. Let $\mathbf{X} = [\mathbf{x}_{1} \cdots \mathbf{x}_{T} ]^T \in \mathbb{C}^{T \times N} $, $\mathbf{y}_i = [{y}_{i1} \cdots {y}_{iT} ]^T \in \mathbb{C}^{T \times 1} $ and $\mathbf{n}_i = [{n}_{i1} \cdots {n}_{iT} ]^T \in \mathbb{C}^{T \times 1} $ be the pilots, receive signal and noise vectors, then, the equation (\ref{H_i}) can be rewritten as:
	\begin{align}\label{Y_i}
		\mathbf{y}_{i} = \mathbf{X}\mathbf{h}_{i} + \mathbf{n}_{i}.
	\end{align}
	
	\subsection{HMIMO Channel Model}\label{S2.1}
	
	According to the derivations in \cite{6,3}, the channel of a HMIMO system can be obtained through Fourier plane-wave series expansion:
	\begin{align}\label{H}
	\mathbf{h}_i=\sum_{l \triangleq\left(l_{x}, l_{y}\right) \in \mathcal{L}} h_{i,l}^{f} \mathbf{a}_l^{f}, 
	\end{align}
	where
	$h_{i,l}^{f}$ is the wavenumber-domain coefficient for the $i$-th user channel, $h_{i,l}^f  \sim \mathcal{C} \mathcal{N}\left(0,\left(\sigma_{i,l}^f\right)^2\right)$.
	$\mathbf{a}_l^{f} \triangleq\mathbf{a}^{f}\left(l_{x}, l_{y}\right) \in \mathbb{C}^{N \times 1}$ represents the Fourier harmonics at the transmitter side,
	\begin{align}\label{a^wd}
	\left[\mathbf{a}^{f}\left(l_x,l_y\right)\right]_n=\exp{\left\{j\left(\frac{2\pi l_x}{L_x}\delta n_x+\frac{2\pi l_y}{L_y}\delta n_y\right)\right\}},
	\end{align}
	where
	$n_x$ and  $n_y$ represent the horizontal and vertical antenna indices respectively, 
	$l \triangleq \left(l_{x}, l_{y}\right) \in \mathcal{L}$ denotes the wavenumber index pairs along the x-axis and y-axis, and $\mathcal{L}$ indicates the elliptical region containing these wavenumber index pairs:
	\begin{align}\label{L}
		\mathcal{L} \triangleq\left\{\left(l_{x}, l_{y}\right) \in \mathbb{Z}^{2} \left\lvert\,\left(\frac{\lambda}{L_{x}} l_{x}\right)^{2}+\left(\frac{\lambda}{L_{y}} l_{y}\right)^{2} \leq 1\right.\right\} .
	\end{align}
	
	Rewriting equation (\ref{H}) in matrix-vector form:
	\begin{align}
	\mathbf{h}_i= \boldsymbol{\Psi} \mathbf{h}_i^{f} ,
	\end{align}
	where
	$\boldsymbol{\Psi}=\frac{1}{\sqrt{N}} \left[\mathbf{a}_1^{f} \cdots \mathbf{a}_L^{f}\right]\in \mathbb{C}^{N \times L}$ represents the wavenumber-domain sparse basis and $\mathbf{h}_i^{f}=[\mathbf{h}_{i,1}^{f} \cdots \mathbf{h}_{i,L}^{f} ]^T  \in \mathbb{C}^{L \times 1} $ denotes the wavenumber-domain channel. The channel dimension $L\approx\frac{\pi L_xL_y}{\lambda^2}$ \cite{3} corresponds to the number of index pairs falling within the elliptical region on the wavenumber plane and is significantly smaller than that of the original channel matrix $\mathbf{h}_i$.
	
	\subsection{Common Support}\label{S2.2}
	
	The literature survey \cite{7,8} indicates that, for multi-user scenarios within a certain communication range, the channel matrices of different users are typically not entirely independent. That is, correlations exist between the channels of different users, especially when users are physically close to each other. 
	Recent studies have also indicated that in multi-user scenarios, the channel DoF of users in the wavenumber domain may overlap \cite{19}.
	As shown in Fig. \ref{fig1}, a group of scatterers in the middle simultaneously contributes to the channels of User 1 and User 2. 
	Such shared scattering clusters physically represent energy concentration within a finite angular range, activating a set of non-zero indices at corresponding positions in the wavenumber plane.
	Consequently, their channel matrices $\mathbf{h}_i^{f}$ may share common support. The introduction of this common support arises from the limited and shared local scattering effects at the base station (BS) side, which is essentially a manifestation of cluster sparsity structure in the multi-user scenarios.
	
	Below, we give the specific expression for the wavenumber domain channel of a user, omitting the user index for brevity. The variance of $h_{i,l}^f$ can be written as:
	\begin{align}
		\left(\sigma_l^f\right)^2 & =\iint_{\Omega_f\left(l_x, l_y\right)} A^2(\theta, \phi) \sin \theta \mathrm{d} \theta \mathrm{d} \phi,
	\end{align}
	where $A^{2}(\theta,\ \phi)$ is the spectrum that depicts the EM field power density,
	\begin{align} 
		A^{2}(\theta,\ \phi)=\sum_{i\in\{1,\ldots,N_{\mathrm{c}}\}}w_{i}p_{i}(\theta,\ \phi),
	\end{align}
	where $w_i$ is the weight of the $i$-th cluster ( $N_c$ clusters in total), satisfying the normalization condition, and $p_{i}(\theta,\ \phi)$ is the probability distribution function (PDF) of the $i$-th cluster, satisfying 3D-von Mises-Fisher (VMF) distribution\cite{3}.
	
	It should be noted that the weight of the $i$-th cluster may vary across users, capturing their different contributions from the same scatterer. Consequently, this leads to distinct energy distributions in the wavenumber domain.
	As expressed in Fig. \ref{fig1}, we can find that the circled common clusters have the same center location, but the size ranges are different. This is where the HMIMO common support set differs from that of mMIMO. In traditional multiuser methods, it is usually assumed that the shared support vectors between users are perfectly overlapping\cite{7,8}, but in HMIMO, we need to consider how to extract the support boundaries of the common clusters.

	Denote $supp(\mathbf{h})$ as the index set of the non-zero entries of vector $\mathbf{\mathbf{h}}$, i.e., $supp(h) = \left\{n : \mathbf{h}(n)\neq0\right\} $. 
	For any user $i$, we define the support set of channel $\mathbf{h}_i^{f}$ as
	\begin{align}
		\Omega_i \triangleq supp\left(\mathbf{h}_i^{f}\right) ,
	\end{align}
	Physically, these indices correspond to the projection of physical scattering clusters onto the wavenumber plane.
	When different users share the same scattering cluster, their common support set is the intersection of each user's support set at the cluster, which is expressed as the maximum shared range of the cluster region in the elliptic plane.
	The common support set $\Omega_c$ can be defined as
	\begin{align}
		\Omega_c = {\cap}_i \Omega_i,
	\end{align}
	Thus, $\Omega_{c}$ corresponds to the set of wavenumber indices where all users exhibit significant channel energy.
	
	In order to simplify the discussion of the system, we assume that all users share the same common clusters\footnote{In practical scenarios, users in close physical proximity typically share common scattering clusters. By incorporating user grouping based on physical location, our assumption can be implemented within the group.}. Then there exists the relationship of the following equation:
	\begin{align}
		|\Omega_c | \geq  s_c, |\Omega_i |\leq s_i, \forall i,
	\end{align}
	that is, there are always shared clusters among multiple users in the scenario, containing at least $s_c$ support vectors, and there is always an upper bound on the support set for each user.
	
	Based on the above support model, the multi-user channel estimation problem can be formulated as a structured sparse recovery problem:
	\begin{align}
		\hat{\mathbf{h}}_i^{f}=\arg \min _{\mathbf{h}_i^{f}}\left\{\left\Vert 	{\mathbf{y}_i}-\bar{\mathbf{X}}{\mathbf{h}_i^{f}} \right\Vert_2^2\right\},\forall i,
	\end{align}
	where  $\bar{\mathbf{X}}=\mathbf{X}\boldsymbol{\Psi}$. 

	\section{Proposed joint estimation algorithm}\label{S3}	
	We propose a multi-user estimation algorithm building upon the graph-cut-based single-user channel estimation method in \cite{5}. The proposed algorithm performs multi-user estimation in an iteratively coupled manner. In each iteration, the algorithm accomplishes three tasks in sequence: 1) Individual support identification, 2) Common support identification, and 3) Individual channel estimation.
	
 	\subsection{Individual Support Identification}\label{S3.1}
	Due to the cluster sparse properties of the channel matrix in the wavenumber domain, the description of this sparse matrix is similar to that of the Markov random field \cite{5}. It models the association between neighboring nodes: when a position in a 2D matrix is a non-zero term, the elements next to it are more likely to be non-zero terms. Undirected graph can visually characterize this relationship.
	 
	For each user $i$, an undirected graph $\mathcal{G}_i$ is constructed to capture cluster sparsity. 
	The vertex set of the graph $\mathcal{G}_i$ is denoted as $\mathcal{V}_i \triangleq \left\{v_l = (l_x,l_y);l \in \mathcal{L}\right\}$,
	and the vector $\mathbf{v}_i \triangleq vec(\mathcal{V}_i) \in \left\{-1,1\right\}^{L\times 1}$ is a binary indicator vector where ``1'' indicates that the corresponding index is a non-zero term, and conversely, ``-1'' indicates a zero term. 
	We introduce a four-neighborhood relation to describe the relationship of vertices in the undirected graph. The edge set is denoted as $\mathcal{E}_i \triangleq \left\{\left\{v_l,v_{l^{\prime}}\right\};|l_x-l_x^{\prime}|+|l_y-l_y^{\prime}|=1\right\}$. 
	
	The channel matrix recovery task can be equated to a problem of undirected graph-cut\cite{5,12}. By finding and cutting the edges connecting the non-zero vertices to the zero vertices, the support vectors for reconstructing the channel matrix can be extracted. 

	Based on the method above, each user individually identifies the support vectors. 
	During the $j$-th iteration of the algorithm, classical residual optimization is applied to extract potential support vectors for different users. The residuals can be calculated as:
	\begin{align}\label{R}
		{\bar{\mathbf{R}}}_i^{(j)}={\mathbf{y_i}}-\bar{\mathbf{X}}{{\mathbf{h}_i^{f}}}^{(j-1)}.
	\end{align}
	The extracted support vectors are then put into a temporary set:
	\begin{align}\label{omega_i}
		\Omega i^{\prime}=\{\lambda(1), \lambda(2), \ldots, \lambda(\widetilde{K})\},
	\end{align}
	where $\{\lambda(1), \lambda(2), \ldots, \lambda(\widetilde{K})\}$ are the indices corresponding to the $\widetilde{K}$ largest values of 
	$\left\|\overline{\boldsymbol{X}}^H( \lambda) \overline{\boldsymbol{R}}_i^{(j)}\right\|_F$, sorted in descending order.
		
	Unlike the OMP algorithm where only the most relevant support vector is extracted, in this step, our algorithm  extracts $\widetilde{K}$ terms which have the largest correlation in the residuals. Extracting multiple candidates allows the algorithm to rapidly cover a potential cluster region within a single iteration and provides a richer set of candidates for the subsequent step. 

	\subsection{Common Support Identification}\label{S3.2}
	Step B is the key joint processing stage where information from all users is aggregated. It takes the candidate support sets $\left\{\Omega_{i}^{\prime}\right\}^K_{i=1}$ from all users (outputs of Step A) and applies joint criteria to distill a high-confidence common support set $\Omega_{c}$.
	In order to extract the common support, it is necessary to filter and prune the support vectors extracted in step A.
	
	Observing the wavenumber-domain channel matrix, it can be found that the value of non-zero terms are always higher in the middle part of the cluster, and decay to the edge until it becomes $0$. 
	The common support set thus captures only indices where all users exhibit sufficiently strong channel energy - typically the central region of the shared cluster.
	Therefore, we set up a threshold condition:
	\begin{align}\label{17}
		{\parallel{\bar{\mathbf{X}}}^H(\lambda){\bar{\mathbf{R}}}_i^{\left(j\right)}\parallel}_F^2>\eta, \lambda \in \Omega_{i}^{\prime}
	\end{align}
	where $\eta$ is a predefined threshold parameter. It ensures that only indices in the high-energy core region of a scattering cluster are considered, preventing weak or noise-corrupted components from entering the common support set. 
	
	After the threshold is satisfied, the process continues to deal with the determination of sharing. As mentioned in Section \ref*{S2.2}, although the users will share the clusters, for each user, the contribution of each cluster is different. Thus, during the operation of extracting the largest $\widetilde{K}$ terms in step A, the vectors inside the clusters with larger weights have a higher probability to be extracted first. 
	This leads to the possibility that a large number of iterations may be required to extract all or the vast majority of support vectors which belongs to all users. It implies that the common clusters we extract can only be utilized in the subsequent few iterations, resulting in relatively limited performance gains. To better leverage the common supports among users, we seek to find the right balance between accurate extraction and performance improvement. Therefore, we have established the following shared decision criteria:
	\begin{align}\label{18}
		\sum_{i}^{} \mathbb{I}(\lambda \in \Omega_{i}^{\prime} )\textgreater K_0,
	\end{align}
	where $\mathbb{I}(\bullet)$ is the indicator function, $\mathbb{I}(\lambda \in A)=1$, if $\lambda \in A$.
	When $K_0$ (where $K_0<K$) users among all users have extracted a specific support vector, that vector is considered highly likely to be a common support vector and is added to the common support set $\Omega_c$. This parameter controls the strictness of common support detection.
	\begin{align}\label{19}
		\Omega_c = \Omega_c \cup \left\{\lambda \right\},
	\end{align}
	
	\subsection{Individual Channel Estimation}\label{S3.3}
	After step B, we add the extracted common support vectors into each user's support set. 
	This operation is where the jointly extracted knowledge is transferred to guide individual estimation. The corresponding vertices $\left\{v_{l};l \in \Omega_c\right\}$ in the undirected graph $\mathcal{G}_i$ of user $i$ take the value of ``1'' .
	\begin{align}\label{24}
	\Omega_i = \Omega_i \cup \Omega_{c}.
	\end{align}	
	These common support vectors reflect the prior information in multi-user scenarios. Through updating the user's individual graph $\mathcal{G}_i$ using the existing support set $\Omega_i$ , inter-user correlations are effectively introduced into the graph, thereby enhancing the precision of individual channel estimation.
	
	We use the graph-cut method  to accomplish this step C. Sparse recovery of the channel matrix in the wavenumber domain can be written as a maximum a posteriori (MAP) estimator by alternately optimizing the support vector $\mathbf{v}_i$ and channel matrix $\mathbf{h}^{f}_i$:
	\begin{align}\label{hf}
		\widehat{\mathbf{h}}^{f}_i=\operatorname{Trim}\left\{\mathbf{0}^{L}\left[:, \bar{v}_{l}=1\right]=\left( \boldsymbol{\Psi}\left[:, \bar{v}_{l}=1\right]\right)^{\dagger} \mathbf{y}_i ; \tilde{K}\right\},
	\end{align}
	\begin{align}\label{v}
		\hat{\mathbf{v}}_i =\min _{\mathbf{v}_i \in\{ \pm 1\}^L}\left\{\sum_{\left(l, l^{\prime}\right) \in \mathcal{E}} V_{l, l^{\prime}}\left(v_l, v_{l^{\prime}}\right)+\sum_{l \in \mathcal{L}} D_l\left(v_l\right)\right\}.
	\end{align}
	where $\operatorname{Trim}$ is the operation of retaining the $\tilde{K}$ largest elements in a vector and setting the other elements to zero, $V_{l, l^{\prime}}\left(v_l, v_{l^{\prime}}\right)$ and $D_l\left(v_l\right)$ is the energy function for edge and vertex \cite{5}.
	
	Estimating individual channel can be transformed into a graph partitioning problem of the undirected graph $\mathcal{G}_i$, which can be solved by the classical alpha-beta-swap algorithm \cite{13}.
	
	The entire multi-user joint estimation algorithm is summarized in Algorithm \ref{alg1}.

	\begin{algorithm}[th!] \label{alg1}
	\caption{Proposed JGC-CE Algorithm.}		
	Initialization: $\widehat{\mathbf{h}}^{f,(0)}_i=\mathbf{0}$, $\mathbf{v}_i^{(0)}=\left\{-1\right\}^L$, the residual $\bar{\mathbf{R}}_i^{(0)}=\mathbf{0}$ and $j=0$  \\
	\KwIn{$\mathbf{X}$, $\mathbf{y}_i$, $\boldsymbol{\Psi}$, $\eta$, $K_0$} 
	\Repeat{ $j \textgreater j_M$ \rm{or} $\bar{\mathbf{R}}^{(j)}$ \rm{converges}}
	{
		$j=j+1$ \;
		\For{$i=1,\cdots,K$}
		{
			update $\bar{\mathbf{R}}_i^{(j)}$ by (\ref{R}) \;
			update and prune $\Omega_{i}^{\prime}$ by (\ref{omega_i}) and (\ref{17}) \;
		}
		update $\Omega_{c}$ by (\ref{18})(\ref{19}) \;
		\For{$i=1,\cdots,K$ }
		{
				update $\Omega_i$ by (\ref{24}) \;
				rebuild $\mathbf{v}_i^{(j-1)}$ using $\Omega_i$ \;
				update $\mathbf{v}_i^{(j)}$ and $\widehat{\mathbf{h}}^{f,(j)}_i$ by (\ref*{v}) and (\ref*{hf}) \;
		}
	}
	\KwOut{$\mathbf{H}^{f,*} = [\widehat{\mathbf{h}}^{f}_1 \cdots \widehat{\mathbf{h}}^{f}_K]$}
\end{algorithm}

    \section{Simulation Results}\label{S4}

	In this section, we numerically evaluate the performance of our proposed algorithm.
	The system parameters are set as follows: $N_x=N_y=65$ antenna elements are packed in a uniform surface array with element space $\delta=\frac{\lambda}{4}$ at the BS side. The system contains five users, with the receiver side configured as a single antenna. The wavenumber-domain channel comprises $N_c=4$ scattering clusters. Two clusters are shared by all users, with the fixed angular position and variable weights. 
	The angular power spectrum of each cluster follows a 3D von Mises-Fisher (VMF) distribution with concentration parameter 	$\alpha=140$. Cluster centers $\left(\theta,\phi \right)$ are drawn from uniform distributions $\mathcal{U}(0,\frac{\pi }{2})$ and $\mathcal{U}(0,2\pi )$, respectively. 
	The pilot matrix $\mathbf{X}$ follows i.i.d. $\mathcal{CN}(0,1)$. Orthogonal pilots are used for different users.
	The parameters in the algorithm are set to be $\eta=0.15$ and $K_0=4$.
	The normalized mean square error (NMSE) is used as the measure of channel estimation performance, given by $\frac{\left\Vert H^f-H^{f,*}	 \right\Vert^2}{\left\Vert 	H^f \right\Vert^2}$. All results are averaged over 500 Monte-Carlo runs.

	Two benchmark schemes are employed for comparison: 1) Applying wavenumber-domain OMP to estimate HMIMO channels individually. 2)The single-user graph-cut-based swap expansion (GCSE) algorithm proposed in \cite{5}.
	
\begin{figure}[t]
		\center{\includegraphics[width=0.45\textwidth]{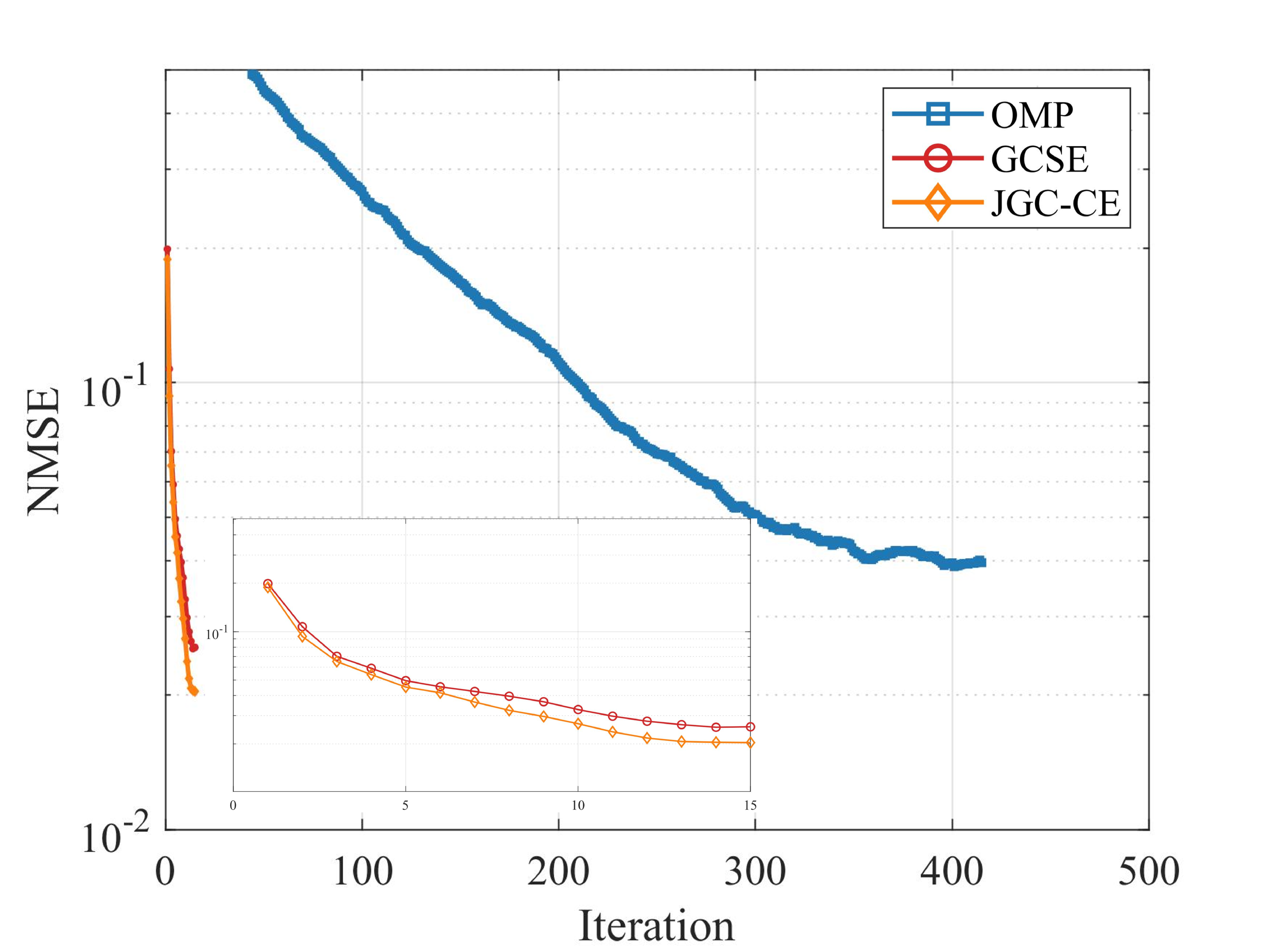}}
		\caption{Algorithm convergence comparison, SNR = 15dB.}
		\label{fig2}
	\end{figure}

	In Fig. \ref{fig2} we examined the convergence of the proposed algorithm.
	It can be seen that GCSE and JGC-CE algorithms can converge after 10-15 iterations, which shows an obvious advantage over the OMP algorithm, which requires 100 iterations or more. The graph-cut method explicitly models the prior knowledge through the undirected graph model \cite{4}. This enables the identification of entire cluster or the majority of points within a single iteration, significantly reducing the number of iterations required. Additionally, the proposed algorithm incorporates information from other users, thereby reducing estimation uncertainty.

\begin{figure}[t]
		\center{\includegraphics[width=0.45\textwidth]{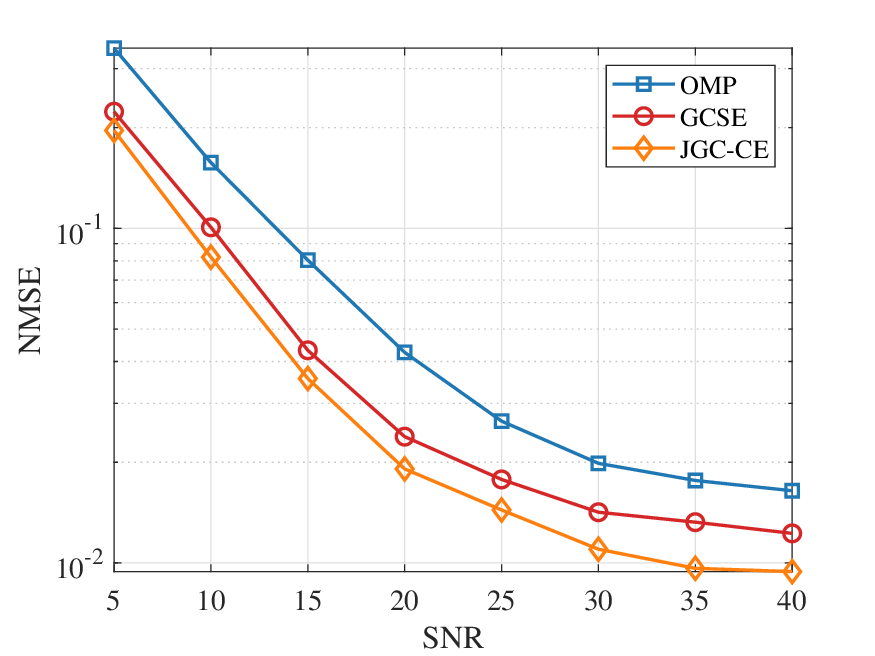}}
		\caption{ NMSE versus SNR under same array aperture.}
		\label{fig3}
	\end{figure}

	From Fig. \ref{fig3}, it can be seen that the proposed multi-user JGC-CE algorithm based on the common support approach improves NMSE performance compared to not using the priori information (GCSE) in all SNR environment. The shared cluster relationship is a characterization of the HMIMO cluster sparsity in the multi-user scenario, and through the extraction of the shared support set, each user can take advantage of the inter-user correlations. 
	This consistent gain is significant as it is achieved with negligible extra computational cost. The core joint operation (Step B) involves only threshold comparisons across users, with complexity $\mathcal{O}(K\widetilde{K})$, which is dominated by the graph-cut step (typically $\mathcal{O}(L^3)$).
	
\begin{figure}[t]
		\center{\includegraphics[width=0.45\textwidth]{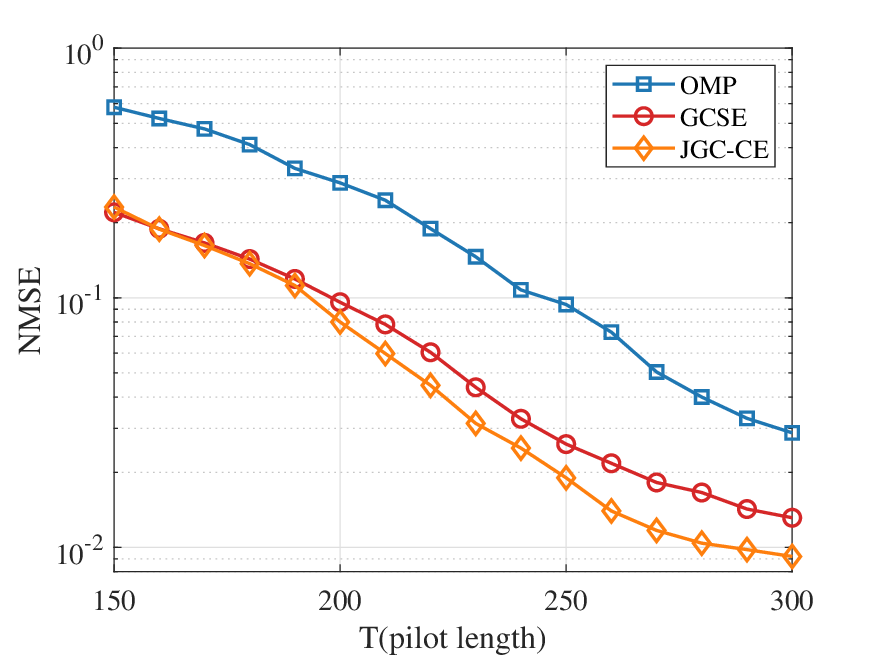}}
		\caption{NMSE versus pilot length, SNR = 30dB.}
		\label{fig4}
	\end{figure}

	Fig. \ref{fig4} plots the estimation performance of the proposed JGC-CE algorithm under different pilot lengths, and it can be observed that the proposed algorithm can reduce the pilot overhead of the multi-user system compared to the traditional estimation method that does not take into account the shared support set. 
	Selecting NMSE = $1.1 \times 10^{-2}$, the pilot overhead is reduced by approximately 11\%. This is because the effective pilot resources used to estimate the common portion can be equivalently reused across multiple users.

	\section{Conclusions}\label{S5}
	In this paper, we have addressed the high-dimensional channel estimation challenge in multi-user HMIMO systems by proposing a novel joint estimation framework that exploits the inherent spatial correlations among users in the wavenumber domain. The proposed JGC-CE algorithm achieves cooperative optimization by alternately updating the globally shared common support set and the user-specific individual support set. The simulation results validates that our framework successfully transforms the inherent multi-user channel correlation into a tangible performance gain.
	This work provides a promising methodological foundation for leveraging user correlations in broader HMIMO scenarios such as the high-mobility scenarios where channels exhibit non-stationary characteristics.

	\bibliographystyle{IEEEtran}
	\bibliography{Refference}

\end{document}